# Studies on Relevance, Ranking and Results Display

J. Gelernter, D. Cao, and J. Carbonell

**Abstract**—This study considers the extent to which users with the same query agree as to what is relevant, and how what is considered relevant may translate into a retrieval algorithm and results display. To combine user perceptions of relevance with algorithm rank and to present results, we created a prototype digital library of scholarly literature. We confine studies to one population of scientists (paleontologists), one domain of scholarly scientific articles (paleo-related), and a prototype system (PaleoLit) that we built for the purpose. Based on the principle that users do not pre-suppose answers to a given query but that they will recognize what they want when they see it, our system uses a rules-based algorithm to cluster results into fuzzy categories with three relevance levels. Our system matches at least 1/3 of our participants' relevancy ratings 87% of the time. Our subsequent usability study found that participants trusted our uncertainty labels but did not value our color-coded horizontal results layout above a standard retrieval list. We posit that users make such judgments in limited time, and that time optimization per task might help explain some of our findings.

## 1 Introduction

Information systems typically list results in order of lexical similarity to the query, with the same word found in query and in retrieved document(s). The goal of much information retrieval research is to improve the quality of the ranked list, i.e., to modify the system such that relevant documents appear higher in the ranked list and non-relevant documents are "pushed" further down [1]. The problem is that relevant documents are not invariably near the top, and when they are, they may be either inadvertently unseen, or simply unwanted because the user has something else in mind. Our three studies with paleontology articles, paleo-trained users and the prototype test how to help create and display results in 'levels of best.'

*What's at the top?*

Retrieval traditionally is measured in terms of recall and precision. Both recall and precision are defined in terms of relevance, with recall being the proportion of relevant documents divided by the total number of relevant documents in a collection, and precision being the proportion of retrieved documents that are relevant divided by the total number of documents retrieved[1].

Scientists disagree as to what relevance means [44]. The etymology is from medieval Latin relevans, past participle of relevare, to lift up. Let's say that an object or action that is relevans somehow relieves a burden. Each person's thought burden—or goal—is different, so what will help one person will not necessarily help another. This presents an immediate quandary in terms of what to present as most relevant, and what should be listed at the retrieval list top.

*How do systems create what's at the top?*

Research is ongoing as to how to create search algorithms which retrieve items that the user is most likely to select. See [34] for example. Relevance levels are a factor of what has been called semantic similarity or proximity. Much research has been done on semantic similarity of result items or per-item sort order, for instance [62] and [61]. The study of how user feedback affects relevance is not the topic of this paper. Here we consider what and how to present results the first time.

Relevance tends to be studied in three separate contexts. What users consider relevant as determined by information-seeking behavior is studied in information science. How systems calculate relevance, whether by machine learning or knowledge engineering-based algorithms and their evaluation, is measured quantitatively in information retrieval. How systems display relevance falls within interface design modeling of human computer interaction. Zhang's Visualization for Information Retrieval, 2008, is among the exceptions in considering within one work both retrieval algorithms and results display.

Why should user relevance, system relevance, and display of relevant items be discussed together, as we do here? We believe that we may add insight to the debate by juxtaposing these aspects.

- *J. Gelernter is with the Language Technologies Institute, School of Computer Science, Carnegie-Mellon University,Pittsburgh, PA 15213.*
- *D. Cao is with the Language Technologies Institute, School of Computer Science, Carnegie-Mellon University,Pittsburgh, PA 15213.*
- *J. Carbonell is with the Language Technologies Institute, School of Computer Science, Carnegie-Mellon University,Pittsburgh, PA 15213.*

[1] National Institute of Standards and Technology, Retrieved Oct 5, 2009 from
http://trec.nist.gov/pubs/trec15/appendices/CE.MEASURES06.pdf

*Subdividing results at the top*

It has been found that many users are uncertain about what they seek at search outset, often because they do not know the possible options [29]. Giving more information about results might aid in the decision.

We could provide more information about results by subdividing by subject, as in the Vivísimo or Carrot2 or iBoogie engines, or in the web search engine by Bernardini et al. [4][2]. Alternately, we could give more information by subdividing results into clusters, subject being only one type of cluster, as in Wolfram Alpha. In Wolfram, for example, a query "piccolo" returns results related to the query by topic, by rhyme, and by etymology, whereas a query on the dinosaur genus "allosaurus" gives information about that dinosaur, a comparison among similar dinosaurs, statistics as to weight, and information about the Saurischia order that contains the genus allosaurus[3].

The result space in our PaleoLit prototype, by contrast, subdivides not by sub-topic, but by three levels of relevance to the query within a single topic, and displays these levels visually.

We try to combine user perceptions of relevance with algorithm design and interface presentation. Our research questions are:

- To what extent do users agree on what is relevant? Or, what should be at the top?
- How should we order the results? Or, how could we determine degree of relevance to the query?
- How should we display possibilities to help users decide what is relevant?

Each study focuses on a related aspect of relevancy. The first study is on what users consider relevant with the data included in the Appendix, the second study is on what and how our system calculates relevance, and the third study considers how the system displays relevancy. Following, we present a time-optimization principle that seems to explain not only why users estimate relevancy the way they do, and why the system that mimics human estimation is successful, but also how the system should display results to facilitate estimating relevancy.

All three studies focus on a group of scientists, a topic domain (even the same set of articles in two cases), and a prototype system we created to support that group and its domain. Our work is with paleontologists, their scholarly literature on paleontology, and our prototype digital library to hold their literature, PaleoLit. Our findings coalesce around the research preferences of a group of specialists and should generalize to other groups of specialists when the domain of the scholarly articles is adjusted. The paper concludes with a summary.

## 2 User Relevance

It has been found experimentally that individuals vary considerably with respect to their relevance judgments [42], and that people are better able to agree on what is not relevant than what is relevant [57]. As a single user proceeds with a search, it has been found that uncertainty gradually decreases [12]. Advanced search interfaces with many field options to allow users to specify queries precisely do not produce documents which users consider relevant to a degree that is statistically significant above search interfaces that are simple [42]. We explain this finding about interface simplicity based on user search uncertainty. That is, there is no point in providing extra interface search options because many people do not have specific search parameters in mind, so the search options would not be used. Our concern is whether, given the same query, those with domain knowledge agree on what is relevant.

### 2.1 Research related to user relevance

Upon what do people base relevance judgments? Some models consider users' cognitive abilities, experience, education, behavior patterns, subject expertise, work tasks, and work environments [8]. Other models focus on relevance of the search algorithm, of the article topic, of the user's idea of relevance, of the user's situation, and his motivation [45]. Still others point to word frequency, the document topic, or the background of the person or situational context [38]. Experiments have found that the nature of the test collection affects such judgments little with the result that the spread of judgments across different collections is comparable [58]. More recently there has been interest in methods for evaluating information retrieval systems without human-generated relevance judgments because creating such judgments requires significant effort and resources [16]

Relevance with respect to content is often measured by means of a 3-level scale [42],[28]. These three levels indicate degree of relevance with respect to the query. The three-level scale used in the Text Retrieval Conference (TREC-7/8) described a highly relevant relevant document discuss the topic theme exhaustively, a mid-level relevant document with topical information but without exhaustive presentation, and a marginally relevant document at the lowest level to only point to the topic. The problem is that even when the scale is well establishhed, users' judgments change. Vakkari and Sormunen relate that half of the TREC documents graded relevant were reassessed upwards as highly, or downwards as fairly relevant or marginal [56]. Interestingly, the TREC interactive track abandoned traditional topic relevance in favor of instance relevance, so that the mention rather than the entire document is graded.

Research in relevance with respect to user judgment considers what aspects of a document influence the selection decision [3]. These have been found to include document traits such as topic, length and date, and source traits such as author and publication. While document traits such as title and abstract seem to influence the selection decision more, it was found that many document characteristics influenced decisions, including article depth, effectiveness, accuracy, clarity, and consensus within the field [3]. These were reflected to some extent by Lee, Theng, Goh and Foo's pilot study [32] that gave users an opportunity to suggest what data they would like to help them make selections. Shiri demonstrated that providing such metadata helps both query formulation and search result presentation [49]. This explains why we designed our PaleoLit system interface displays title, abstract and date to help users make decisions, desribed below in the third study.

[2] On October 26, 2009, Vivísimo is located at http://vivisimo.com; Carrot is at http://www.carrot2.org; iBoogie at http://www.iboogie.com.
[3] On October 26, 2009, http://www.wolframalpha.com

We will use the relevance judgment of these participants as a benchmark to evaluate the retrieval of our system in the second study. But relevance judgments are known to differ across judges and for the same judge at different times [48]. Critics question how valid conclusions can be drawn when the process is based on something as "volatile" as relevance [58].

Our research questioned the extent to which domain experts agree on relevance. To this end, we used a document sample of 30 and allowed flexible time to make decisions. A larger document sample would have been impractical since the task, although participants were paid, would have become a chore. We had asked participants to qualify their judgments and explain why they made their choices, but we received little by way of response beyond the categories each chose.

## 2.2 User relevance study

*Objective*

We wish to determine the extent to which those knowledgeable on a subject domain agree with respect to per-item degree of importance.

*Participants*

Our participants were an undergraduate who had taken some paleo-courses, a graduate student in paleontology, and a professional paleontologist who is a museum curator.

*Data sample*

Our collection goal was to find 30 digital articles in journals in paleontology, paleo-biology and geology on one animal and one plant topic. Bibliographic citations to each of the articles are included in the Appendix for the purposes of study replication. We use "ginkgo" as the plant (a living species that has survived since the Mesozoic era), and "allosaurus" a carnivorous dinosaur genus thought to be somewhat earlier than Tyrannosaurus Rex. We collected about two-thirds of the articles by keyword search on our selected plant and animal, and then downloaded some articles randomly from paleo-journals to complete the quota of 30. These randomly-selected articles had neither search term and pertained to neither category.

TABLE 1
RELEVANCE JUDGMENTS OF THREE PARTICIPANTS CONCERNING 30 PALEO-RELATED ARTICLES (BIBLIOGRAPHIC CITATIONS FOR EACH ARTICLE ARE LISTED IN THE APPENDIX)

| Ginkgo H | Ginkgo M | Ginkgo L | Allosaurus H | Allosaurus M | Allosaurus L | Not relevant to Ginkgo or Allosaurus |
|---|---|---|---|---|---|---|
| 1,1,1 | 15 | 5 | 8,8,8 | 25 | 6,6,6 | 2,2,2 |
| 3,3,3 | 22 | 11,11 | 25,25 | 16 | 9,9 | 4,4,4 |
| 18,18,18 | 27 | 27,27 | 13 | 24 | 12,12 | 7,7,7 |
| 11 | 29,29 | 29 | 16 | 13 | 13 | 10,10,10 |
| 15,15 | 5 | 28,28 | 19,19 | 9 | 16 | 14,14 |
| 21 | 14 | | 24 | 12 | 19 | 17,17,17 |
| 5 | | | 26 | 23 | 23,23 | 20,20,20 |
| 22 | | | | | 24 | 21,21 |
| | | | | | 26,26 | 28 |
| | | | | | | 30,30,30 |

*Measurement*

We counted per-article agreement among participants. Based on others' findings, we knew that 3/3 consensus would be found rarely, so we graded full agreement as 2/3 or 3/3, and partial agreement as 1/3.

*Procedure*

We gave our participants two search queries–ginkgo and allosaurus–and asked them to judge whether the given articles were of high, medium, low or no relevance to the queries. We tried to make the task pleasant and easy, so we gave no time limit, and asked the participants simply to sort the articles into 7 piles: allosaurus, high, medium and low relevance; ginkgo high, medium and low relevance; or about neither allosaurus nor ginkgo. We did not provide direction as to what constituted each relevance level so that we could compare judgments. Instead, we asked each participant to write criteria by which the relevance levels were judged. Only one of the three did what we asked (Table 2).

*Findings*

Results are in Table 1.[4] We confirmed that participants are more likely to agree on what is not relevant than on what is relevant. When they did agree upon what is relevant, they rarely agreed on relevance to the same degree. Both agreement on irrelevancies and non-agreement on relevancies is consonant with findings in Text Informative Retrieval Conference (TREC) evaluations [18] [7].

How participants ranked the 30 articles:

[4] These results served as a benchmark for scoring system relevance in in the second study, which used the same article sample as the first study.

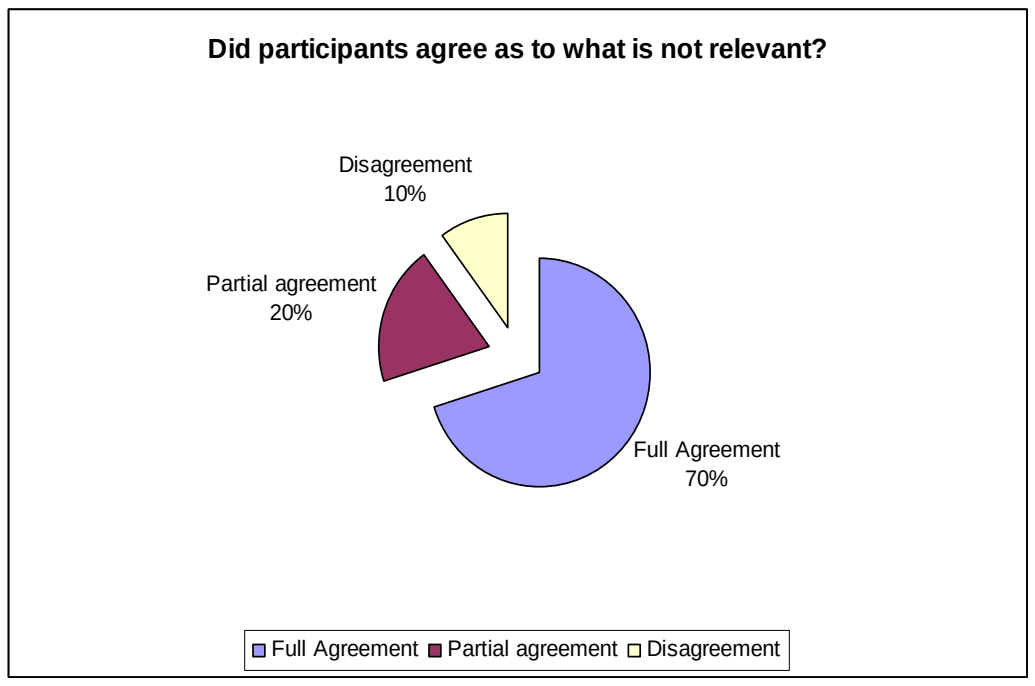

Fig. 1. Chart showing agreement as to what is not relevant.

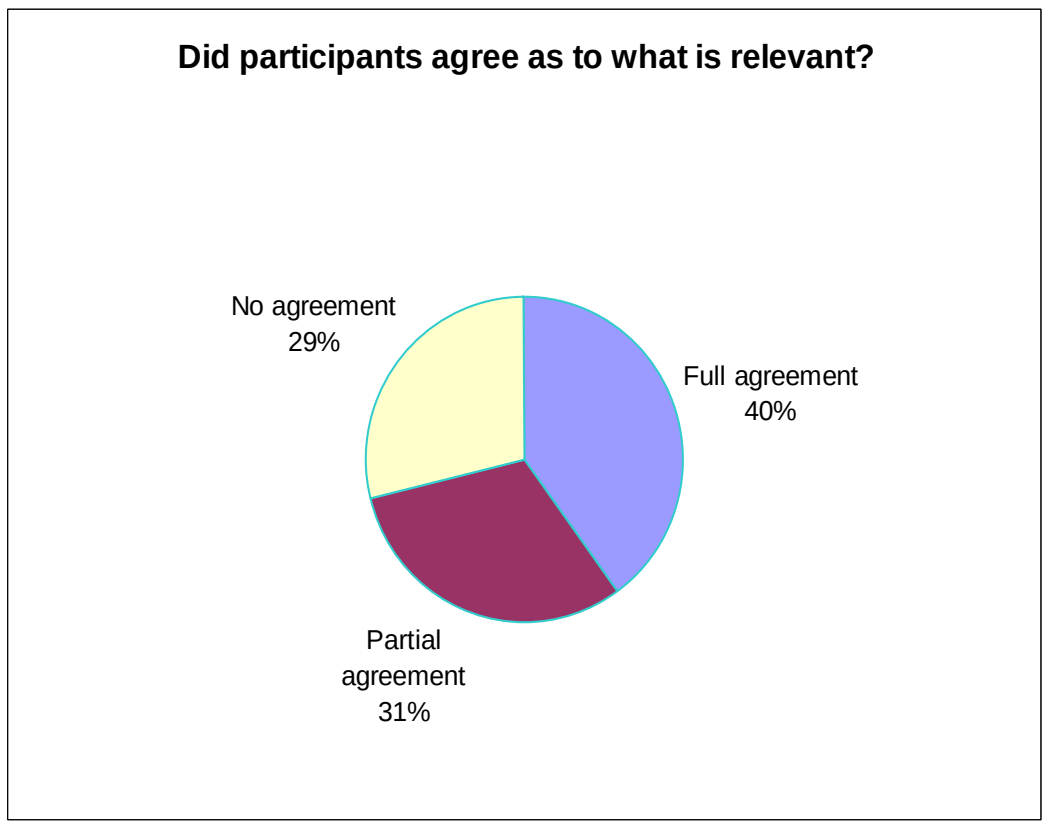

Fig. 2. Chart showing agreement as to what is relevant

TABLE 2
PARTICIPANT DEFINITION AS TO WHAT CONSTITUTES HIGH (VERY RELEVANT), MEDIUM (RELEVANT) AND LOW (SOMEWHAT RELEVANT) ARTICLES. "SUBJECT" HERE REFERENCE TO THE QUERY TERM

| Very relevant | Relevant | Somewhat relevant | Other |
|---|---|---|---|
| Subject mentioned in title and/or abstract | Subject as none of "Very relevant" categories but is coded for phylogenetic analysis and paper appendix data | Subject mentioned multiple times, even in title, but central question is of little interest to paleontology | Subject is not mentioned anywhere in article, incl. its figures or appendices |
| Subject not in title/abstract but multiple times in paper in the context of the paper's central subject | Subject specimen occurrence and/or locality data mentioned once or twice | Subject mentioned multiple times, even in title, but no new information or interpretation relevant to subject is provided | Subject mentioned once or twice but only in passing with no new/useful information about subject provided |
| Even though subject not in title/abstract, specimens of the subject are in illustrations | Subject mentioned multiple times outside of title/abstract and used as comparison to main point of paper; aids understanding of subject | | |

Table 1 shows that the participants agreed least as to the middle relevance level. As the in-between condition, we have confirmed Lesk and Salton [33] in that disagreements tend to affect the borderline documents.

Figures 1 and 2 demonstrate the extent of participant agreement. Fig. 1 shows that they agreed as to the not-relevant classification 70% of the time, while Fig. 2 shows that they agreed on what was relevant only 40% of the time.

What constitutes relevance? Our participants clearly disagree. Only one participant answered our request to report criteria used in making the decision as to whether an article is relevant. The given criteria concerned bibliographic fields and frequency of term occurrence, as shown in Table 2.

While the participant focused on the query subject in defining document relevance, the location of mention of the query term was clearly pivotal in the relevance assessment (Table 2).

*Limitations*

We would have been able to draw conclusions with greater certainty had we more participants in each group, say 3 undergraduates, 3 graduate students and 3 professional paleontologists, rather than 3 altogether. Further testing could include more participants. We had assumed that each participant's knowledge level roughly corresponded to the number of years he or she spent in the discipline. However, we could have asked each participant for a self-rating as to acquaintance with research on "ginkgo" and "allosaurus".

## 2.3 Contribution of our user relevance study

We have validated within a specialist domain what others have found more generally: that users do not agree as to what is relevant (only 40% of the time), and that they tend more to agree as to what is not relevant (up to 70% of the time).

Individual differences may be called upon to explain differences in individual decisions, but for understanding behavior patterns we turn to the social sciences. Psychologist-economist Herbert Simon believed that differences in decisions usually stem from differences in knowledge, where knowledge includes both skill and information [51]. We take Simon's assertion one step further, some people without knowledge or skill care less and so might spend less time and make decisions faster. We did not measure the time it took each participant to complete this exercise, but we should test this assumption in the future.

We might use this result to create guidelines that would improve system design. Developers could alter the sorts of choices retrieved by the system, the sort of interface presented, or both. For example, we could alter the choices retrieved by re-balancing results toward either high recall or high precision, as suggested by Si and Callan [50]. The system might lean toward high recall for a more knowledgeable user, or high precision for a quicker and wider search. Otherwise, we might help people make decisions faster by providing bibliographic information that reveals content relevance (article title, date, abstract), and indicating which articles are more relevant (our relevance labels). These latter suggestions were implemented in our prototype and evaluated by users, as described in the third study.

# 3 RELEVANCE CALCULATED BY THE SYSTEM

Here we introduce a paradox. "Better" search engines are deemed to have a higher average relevance, which amounts to a balance between recall and precision. The problem is that this average is calculated by considering what is and what is not relevant. But it has been found (as mentioned in the context of the first study) that there is more agreement on what is not relevant than what is, where we use participants' rankings of 30 articles as a standard to measure system relevance rankings. The purpose of this section, then, is to show that our retrieval algorithm that uses location of words in an article ranks results closer to how users order results than does a standard bag-of-words system.

How closely do standard system and user relevance rankings match? Patil, Alpert, Karat and Wolf [39] created a tool to compare the relevance ranking provided by a search engine to that judged by users. Another experiment with a basic ranking algorithm (number of times the word is found, etc., with no ontology) found that only 12% of the items presented as highly relevant were also ranked as highly relevant by 10 participating users [38]. The authors conclude that "this implies that pure algorithmic ranking does a very bad job at providing relevant results to the user in the top 10 positions of the result list."

Perhaps users have a sense that their own estimation of what is relevant does not necessarily coincide with a search engine's retrieval. That would explain why few use Google's "I'm feeing Lucky" button that displays a single result for the given query. This button has accounted for 1% of all searches.[5] Alternatively, it is possible that the "Lucky" button is underused because users do know what the feeling is all about and do not wish to experiment. However, it is also possible that user prefer a selection of choices. Our hypothesis could be tested with large-scale survey of those have used or ignore the "Lucky" button.

### 3.1 Research related to system relevance

Retrieval relevance is calculated mainly through one of three models: Boolean, vector and probabilistic [36]. The Boolean model represents documents as either relevant or not. The vector model represents documents as a set of attributes, or vectors. The probabilistic model represents documents according to their probability of being relevant, and that relevance is generally calculated with respect to the document collection as a whole. Only vector and probabilistic models offer relevance to a degree than the binary relevant or not. Our model has properties of a vector model in terms of per-article weighting.

We constructed our algorithm by means of document representation cues, ontologies, and corresponding weighting. A brief look at what others have done should clarify how our work compares.

To extract metadata from an article in Adobe Portable Document Format (PDF), some have performed rules-based extraction [23], while others have used image-based extraction in which pre-defined bounded boxes indicate placement for different fields [26]. Han et al. [23] have had very good results generating metadata using machine learning techniques. We were unable to run their open source HeaderParse algorithm that comes from the CiteSeer suite. CiteSeer is a digital library of scientific literature and a search engine which uses heuristics to distinguish title from abstract from citations [6]. Since our work on this area of the research has concluded, an easy-to-use pdf to doc converter has become freely available[6], although for our work we used the open source pdf2xml from Sourceforge. For our purposes, we generated heuristics for title, abstract, full text, caption, and references through examination of the training data.

To determine which words in a document are most suitable for indexing requires reducing the dimensionality of the data. We followed some standard pre-processing procedures to reduce the number of words. For example, we used stop word lists that have been shown to reduce noise [19]. To improve matching, we stemmed words [20], matches phrases (or n-grams) as well as single words [54], and expanded queries via ontologies [60].

Relative ranking among documents is typically determined by tf-idf weight (term frequency–inverse document frequency) to at least some degree. A term is any content-bearing word. The tf-idf incorporates frequency of term occurrence in a document, length of the document, distribution of the term throughout the document collection (inverse document frequency), and the distribution of the term within the document [44]. We do not use this formula for ranking for two reasons. We do not find document length a useful determinant because there is little variation among length in our set of journal articles. Also, we do not use distribution of the term in the collection because we expect the nature of the collection to vary considerably over time. We do use the location of term within the document to help determine of relevance. Also we use ontologies to rank retrieved items, as has been done before [15].

Our weighting formula classifies results with respect to uncertainty levels. We determined level of uncertaintly by induction, looking at a sample to determine where in the document the query term could be found. Alternatively, we could have used a scoring continuum with an experimentally-derived threshold score for each uncertaintly level would work just as well, with the threshold determined machine learning [67]. We did not have a large enough document sample, however, to determinee what optimal threshold scores should be.

### 3.2 Our PaleoLit program

We created heuristics by studying a wide variety of paleontology and paleobiology publications. Our 116-article training set contains 28 publications which feature paleoanimals and plants, including major titles such as Palaois; Journal of Vertebrate Paleontology; Evolution; Journal of Human Evolution, Historical Biology; Paleobiology; Paleogeography, Paleoclimatology, Paleoecology, and the American Journal of Botany. These and similar journals formed the sample from which our rule-based algorithm

[5] http://en.wikipedia.org/wiki/Google_search#.22I.27m_Feeling_Lucky.22

[6] Solid Converter pdf to doc download, Retrieved May 22, 2010 from http://www.prime-download.com/search-results.html?query=Solid%20Converter%20PDF%204.0&aid=9330000

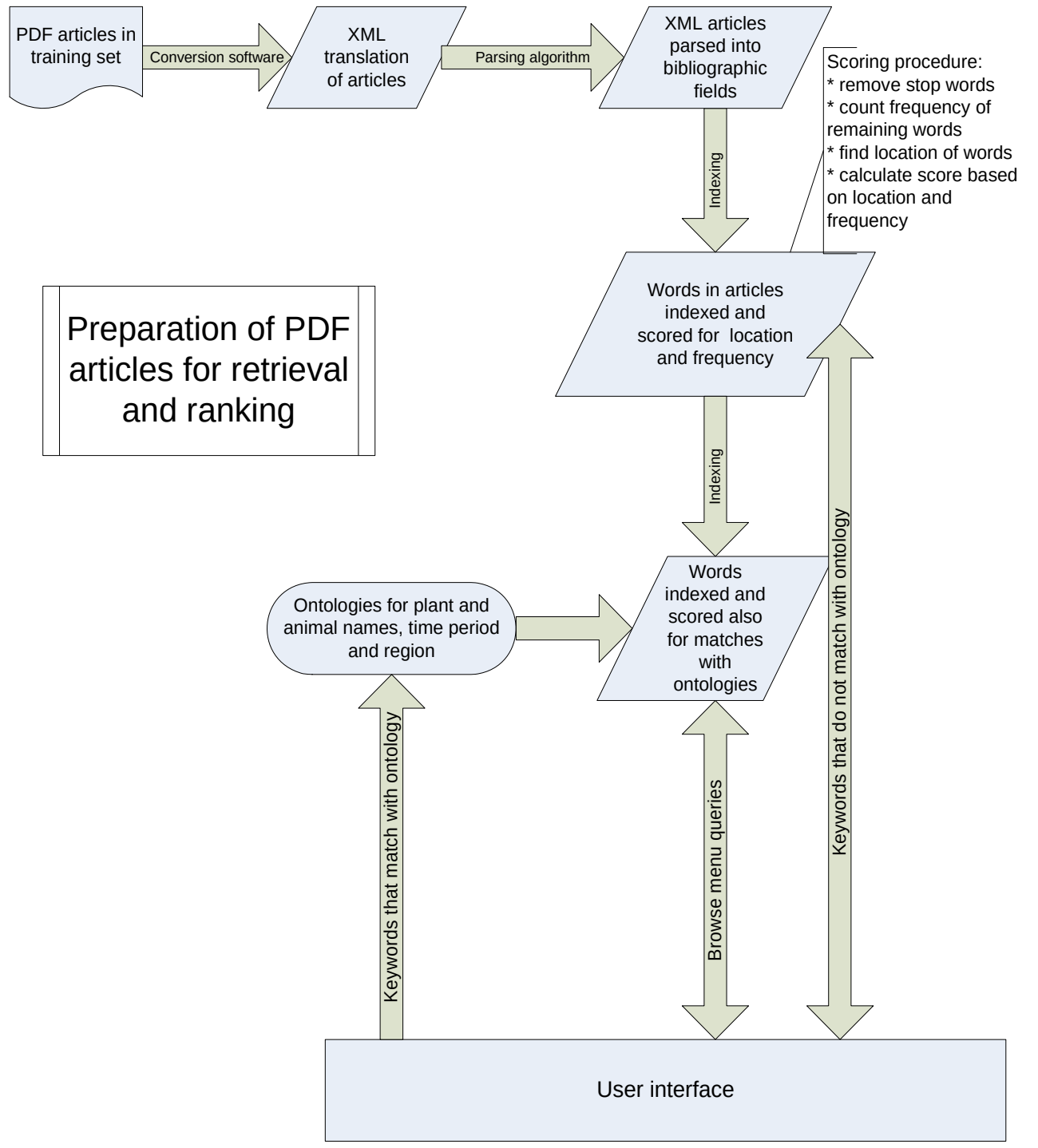

Fig. 3. Chart showing preparation of PDF articles for PaleoLit browse and keyword search and retrieval.

was constructed.

Our program is described in some detail in Gelernter et al. [22]. Here we give an overview (Fig. 3). During pre-processing, documents are converted from pdf to xml. Then metadata in the form of bibliographic fields such as title and abstract are generated from the xml. We remove stop words, and index the remaining words and phrases as found in our three ontologies: the downloadable version of the Linnean classification system from the Index of Organism Names (ION), the standard geologic time scale, and a set of world countries and cities for region[7]. Because our prototype is a beta test site accessible via the Geosciences network (GEON) portal,[8] we used the GEON ontologies for time period and region.

## 3.3 Weighting model

We do use frequency and distribution of term in the document. We add subtlety to the number of terms in a document by considering not only how many instances of a term there are, but where in the document they occur (title, abstract, illustration captions, and so forth). Similarity to the query or relative weight of the term is determined in part by the bibliographic zone in which a match is found (title, abstract, full text, references, etc.). This we call level of fuzziness. We use levels: high, medium and low. For users, in system testing and in the interface, we translate these levels as "Highly relevant", "Relevant", and "Somewhat relevant".

We experimented with how different weights in bibliographic fields multiplied by the number of term occurrences would affect item ranking. Our weighting formula appears below for the sake of replicabilty. The cross-outs[9] show how we changed weighting after experimentation with the training set.

Match in title ~~10 20~~[8] 12 x number of occurrences
Abstract 8 ~~17~~ 10x number of occurrences
Author-supplied keywords ~~10 20~~ 12 x number of occurrences
Caption ~~7~~ 8 x number of occurrences (unless there are two or more occurrences of organism names in the *same* caption; then only 3 x number of occurrences
Ersatz abstract ~~6~~ ~~10~~ 9 x number occurrences
Full text, in words 1-2000 4 x number of occurrences
Full text, in words 2000- end (excluding references)
2 x number of occurrences

Length of term match with ontology
Exact match. Weights for query term matches with the ontology
2 words exact match (ex. Water lily, or Upper Paleozoic)
10 x number of occurrences
1 word match (ex. Lily) 5 x number of occurrences
1 word match on child 3 x number of occurrences
1 word match on father 2 x number of occurrences

Context of term
If two terms from different indexes (name+time) or (time+region) or (region+name) appear in the same sentence, x 5

We first set weights to make a match in the document title to score 10, match in the abstract 8, match in author-supplied keywords 10, captions 7, and the first paragraph of the article 6. Initial testing showed these weights too low. We then raised the match in title to score 20, in the abstract 17, 20 in the author-supplied keywords, and so forth. Testing proved these weights too high. We settled on 12 for the title, 10 for the abstract, 12 for the author-supplied keywords, and so on, with all weights specified in the table below.

We determined by looking at a large number of documents that document sentences that mention terms in two or more of our three designated indexes (region, time and subject) is an important sentence, and therefore these terms should score higher. We designated these as score of 5.

Our weighting formula extends to matches with the ontology, again multiplied by the number of occurrences. A phrase match with the ontology, at 10, scores higher than a single word match with the ontology, at 5. A match on the lower element, or child, in the hierarchy, which we scored as 3, is slightly more relevant than a match higher in the hierarchy, or parent, which we scored as 2.

The chart below illustrates how the ontology influences weighting by augmenting the query. We use the Linnean classification system (Phylum, Class[10], Order, Family, Genus, Species) for organism names. Both our sample queries use "allosaurus" genus for organism name, so we discuss this in detail. Suppose the user enters "allosaurus". (See Fig. 4) The query matches all instances of "allosaurus" as found in the documents. "Highly relevant" documents retrieved mention allosaurus in the article, and less relevant documents mention a species *Allosaurus fragilis* or *Allosaurus tendagurensis*.[11]

[7] Our sample from Trilobita came from the Reuters Index of Organism Names http://www.organismnames.com/, supplied thanks to Reuters; our taxonomies for gingko and a few other species were downloaded from the Paleobiology Database, http://www.paleodb.org/cgi-bin/bridge.pl.
[8] http://www.geongrid.org

[9] The cross-outs from our testing are preserved to show how we have adjusted our weighting system.
[10] The Linnean category called Class is not to be confused with the categories our system uses for classification.
[11] A limitation of our weighting system is manifest when an organism is mentioned in more than one way in the same sentence, and each men-

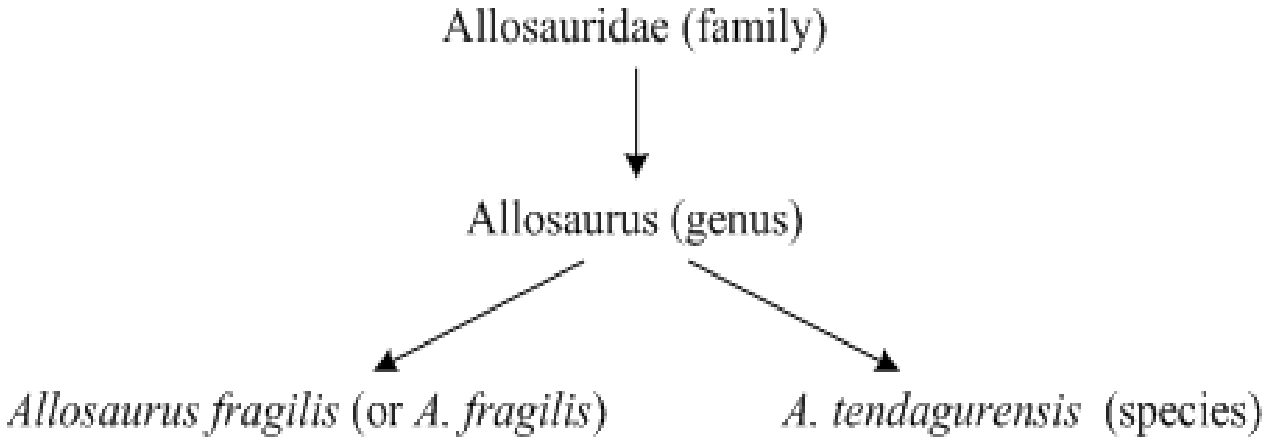

Fig. 4. Organism classification according to the Linnean system.The species name customarily is written either repeating the genus or with the genus abbreviated.

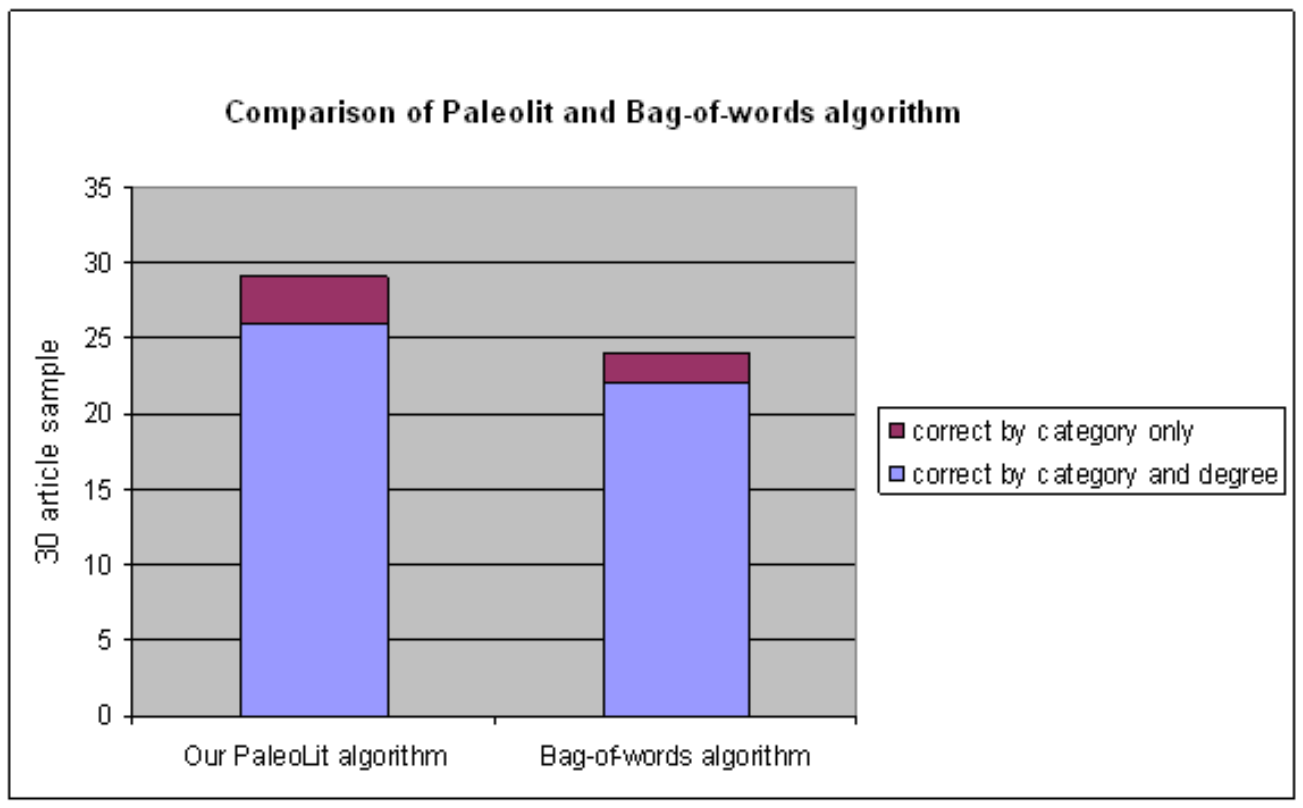

Fig. 5. Chart comparing results of our algorithm to that of an algorithm that does not consider location of term in the document, with the benchmark determined by those with domain knowledge

## 3.4 Study to test system relevance

*Objective*

How well do rankings produced by the PaleoLit algorithm match user-defined rankings for the same articles?

*Data sample*

The effectiveness of the algorithms depends also upon training set quality, where quality measures representativeness of the ultimate collection. We used the same 30 article sample with articles on "ginkgo" and "allosaurus" as for the first study.

*Measurement*

We used for a benchmark the specialist-ranked 30 paleo-article gathered in the first study. We compared results of our algorithm to those of an algorithm that does not account for location of words in articles but only word frequency.

*Procedure*

Search algorithms are typically evaluated according to mathematical formulae that involve recall and precision. But these terms are themselves defined in terms of relevance, itself elusive and variable from person to person. So we evaluate the system by comparing how the algorithmic ranking compares to manual ranking for the same items.

We created a bag-of-words algorithm that used ontology categories for organism name classification, and the above criteria for category degree to run through the 30-article test sets. Everything about the plain algorithm was the same as ours except its lack of weighting by area in the document in which the word was found. For degree of relevance, the plain algorithm assigned an item as Low relevance if the word was found twice, Mid if it was found three or four times, and High if it was found 5 or more times. The evaluation consisted of parsing the 30-article test set, and then running the test set through the plain algorithm as well as through our algorithm that reflects document representation, or place in the document where the match term is found.

*Findings*

Our algorithm that used location of term in the document as well as term frequency was 87% accurate with respect to relevance level, as opposed to the bag-of-words algorithm that was only 73% accurate with respect to relevance level (Fig 5).

*Discussion*

Recall that our research question was 'How should we order the results?' Our study compared our system's relevance ranking to that of a system that did not use location of term in document for the ranking, to that of human ranking. Our PaleoLit system did quite well in comparison to the rating from Ochoa and Dubal [38] in which the system relevance agreed with users only 12% of the time. For the 30 article set, we had 80% agreement with at least one domain specialist. That is, for queries "ginkgo" and "allosaurus", High, Medium and Low relevance ranking assigned by our PaleoLit system agreed with the three participants' rankings (from the first study that created our benchmark) at least 80% of the time. The comparison to the plain retrieval system shows that some of the success of the PaleoLit ranking is attributable to the fact that it uses location of a term in the document to assign ranking.

## 3.5 Contribution of the system relevance study

We present a prototype search system that has been shown to rank results more similarly to how domain specialists order results than does a bag-of-words retrieval algorithm. One of the features of our system that improves ranking is our use of document representation, that is, the field the term is found is, for weighting.

But our algorithm is not tailored to each user individually. Tailoring requires personal measurements that are rarely volunteered, and if taken without user notice, tion is counted separately rather than as one mention. So, for example, "conodont Hindeodus parvus" counts as three in our system for class conodont (conodonta), genus and species Hindeodus parvus even though it is actually a reference to a particular creature. We retain this inconsistency for the practical reason that scanning the text more than once to correct this would slow the indexing and match process enormously.

are subject to privacy infringement. This paper does not venture into the domain of the personal. However, we might be able to get information as to a searcher's level of skill or interest in the material if he would volunteer this openly. Results could then be skewed toward the recall or precision end of the retrieval spectrum accordingly.

## 4 Relevance in Interface Design

How might search results be presented to help users select what is relevant? We introduced two features into our prototype PaleoLit system: relevance labels and color blocks. The horizontal display of the color blocks increases the number of per-screen results shown.

Presently, standard results display in a list. But retrieved items at the top of the list may be of high or low relevance with respect to the query. Results may be low relevance to the query even though at the top of the result list if they are simply more relevant than everything else in the database. It is in the self-interest neither of search engine companies nor of merchandise sellers to point out to users that top listed results are not particular relevant to the query, since search engines and sellers might profit from user selection of choices provided rather if the user selected another search engine. Our research, by contrast, is in the interest not of the data owners, but instead of the data consumers.

We aim to add clarity to the result display and save users' time by making it obvious when top-listed results are not too relevant. We do this by actually labeling results as "Highly relevant," "Relevant," or "Somewhat relevant". Even without labels, an experiment showed that users assume results at the top corner of the screen are most relevant [2]. But another experiment found that searchers did not read 30-50% of the relevant items in the top ten positions [55]. Might labels draw users to relevant results? We aim to create a visualization that "facilite[s] the human cognitive process" [66] and aids result selection.

### 4.1 Research related to relevance display

Standard search results list according to a single dominant characteristic, with the most relevant at the top. But there are other options for result display as mentioned in this paper's introduction. Results may be clustered by sub-topic, such as the dynamic rainbow of the Carrot clustering engine, or by a variety of relevancies as in Wolfram Alpha. To demonstrate relevance, different colors are used less often than different shades of a single color. Compare this to geographic maps in which the darker shades indicate 'more' of some theme (rainfall, population, etc.) and the lighter indicate less of that theme. Blurring also has been used as a metaphor for uncertainty [40]. It is not our intent to catalog visualization types, as has been done admirably by Chen [9], or to discuss visualizations found in digital libraries [5][21]. Here, we focus on visualizations of levels of relevance or uncertainty with respect to the query.

It was found in at least one user study that although query options were important, searchers were more concerned with the option's usability than with option particulars [31]. Rather than use icons that might be hard to understand, we used color shades to indicate relevance level. The horizontal display also allowed us to include more results per screen, requiring the user fewer clicks to view more results. Thus, our other answer to improve usability in our prototype was to display by color grid.

A grid is a standard alternative to a list in many commercial softwares. See the Smithsonian Research Information Systems, for example, with list and grid display options[12]. The EBSCO corporation makes a vast collection of journal articles available for online search. EBSCO's visual search option also displays by blocks. A grid often is used for thumbnail images, when showing more results per page decreases wait time associated with loading image-laden pages. It is, however, more novel in bibliographic results listing, when the grid fills not with pictures but with citations.

We do not question grid effectiveness as in Allan et al. [1], because we wanted to know whether users would turn to the grid in the first place. If they would not, the idea of grid effectiveness is academic. Users might adapt to and actually prefer the grid upon gaining familiarity, but we tested for immediate impressions.

### 4.2 Study

Our PaleoLit digital library prototype of scholarly articles on paleontology was the basis for this study. Its target user base is professional paleontologists. We launched the first version in time for the 2009 North American Paleontological Conference. With the permission of the conference chair, our questionnaire was included in the packet distributed to conference attendees.

We elicited user opinions in the form of a survey.[13] We hope to use these results to improve the design of our prototype. User-centered design employs "active involvement of users for a clear understanding of user and task requirements, and the iteration of design and evaluation" [35]. Moreover, user-centered design can improve system usability [59].

We offered results display in two formats: list view, and a horizontal grid where relevance is shown by color blocks. Title, date and abstract are provided for each article in both list and grid view.

*Objective*

We sought feedback from our intended user group as to how we had designed the relevance display.

*Participants and sample size*

Our participants were paleontologists who attended the June 2009 North American Paleontological Convention.

[12] Smithsonian Institution Research Information Systems, at http://collections.si.edu/search/results.jsp?kiowa&view=grid&start=0
[13] The entire survey can be viewed at http://paleosearch.rutgers.edu/paleo/popup.jsp . An earlier version that shows the bright purple is at http://paleosearch.rutgers.edu/paleo/index.html , available on the web as of June 3, 2010.

Participants were self-selected. 17 responded.

*Measurement*

Participants responded using Likert-scale selections of never—rarely—frequently—always. We did not provide a neutral scale choice because it gives little information.

*Procedure*

We ran a pilot test among those with paleo-training to be sure that the questions were non-ambiguous and to gain insight by talking to users. The questionnaire we distributed at the conference was a one-page survey on user preferences about the prototype design. We used separate questions to ask the result labels and the grid view. Specific questions included:

Would you use our result labels Highly Relevant, Relevant and Somewhat Relevant

☐ Never ☐ Rarely ☐ Frequently ☐ *Always*

b) if you answered (a) as Never or Rarely, should we change the labels or list results some other way? How?

________________________________________

Would you use PaleoLit Search grid view

☐ Never ☐ Rarely ☐ Frequently ☐ *Always*

Participants based their answers on online examination of results displayed in one of two PaleoLit screens, similar to those shown Figures 6 and 7.

*Findings*

Of the 17 who responded, 58.8% thought our relevance labels were frequently or always valuable, but only 35.7% would ever use the grid view. How might we explain views on the labels? A look at feedback from our pilot test helps, because we received little on the actual survey. One wrote about the value of the PaleoLit relevance labels "I feel it [labeling] is marginal because the relevance of the article seems subject to the individual…". Two recognized that adjusting label relevance rankings to user ranking is part of the research. "I think it could potentially be valuable," wrote another, "but right now it's kind of confusing how it ranks the relevance, since sometimes articles I would consider most relevant show up in 'somewhat relevant'." Another wrote that the labels are "marginally useful at present. Although I like the idea of rating articles by relevance, at present the implementation seems to be somewhat imperfect."

*Discussion*

There was no correlation between those who responded that they liked the relevance labels and those who liked the horizontal results layout. Perhaps the horizontal grid was unfamiliar. It has been found that the success of visualization techniques depends upon user familiarity [40]. On the other hand, it is possible that users did not dislike the grid but simply preferred the familiar list display. Alternatively, it might have been due to our unfortunate choice of overly bright shades of purple and violet that turned people away from the grid. The fact that the grid places many more results on the screen than the list (12 in grid vs. 6 in the list) was not commented on by any participant.

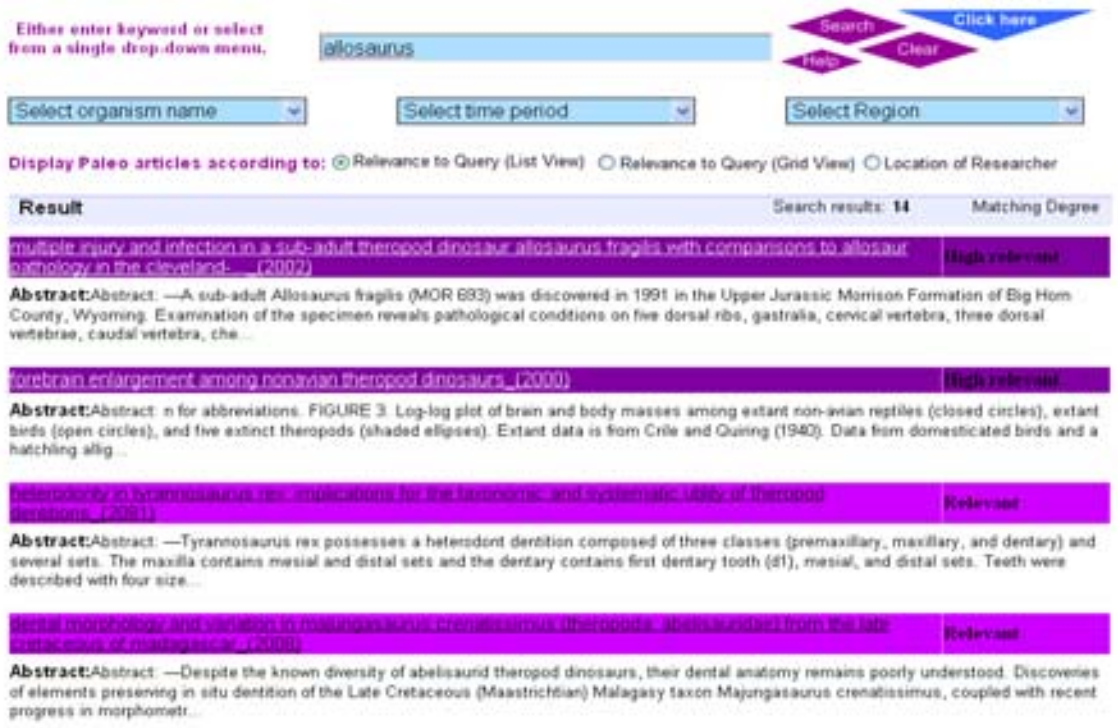

Fig. 6. PaleoLit list view search on "allosaurus". A live version of this prototype is at http://paleosearch.rutgers.edu/paleo/index.jsp.

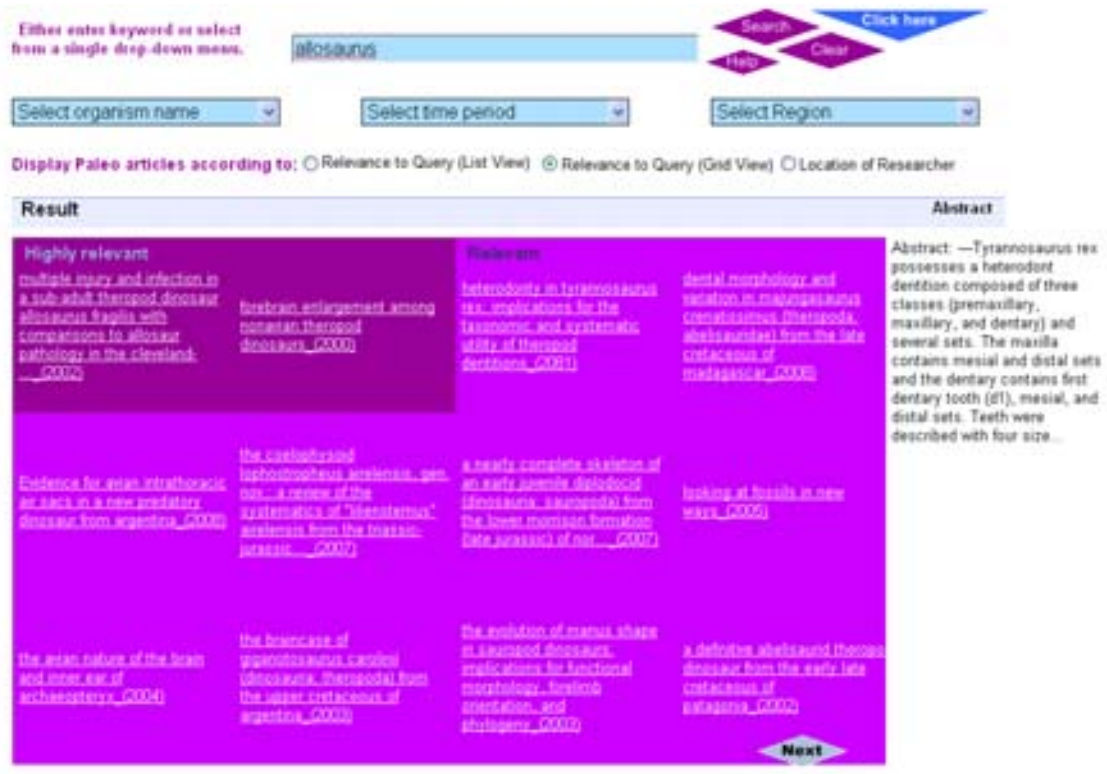

Fig. 7. PaleoLit grid view responses to search on "allosaurus," with blocks in shades of the same color labeled "Highly relevant", "Relevant" or "Somewhat relevant".

## 4.3 Contribution of relevance display study

Many of the participants liked the idea of the relevancy labels, although we must still try to improve the harmonization of system high relevance to user high relevance to the extent possible. The horizontal layout was not met with enthusiasm, perhaps due to its unfamiliarity, or a dislike of overly loud shades of purple, or perhaps simply because a standard result list was thought preferable. In terms of user-centered design, we replaced the bright purples to cooler blues in the next version of the interface. In addition, we added uncertainty labels to the list view (Fig. 8).

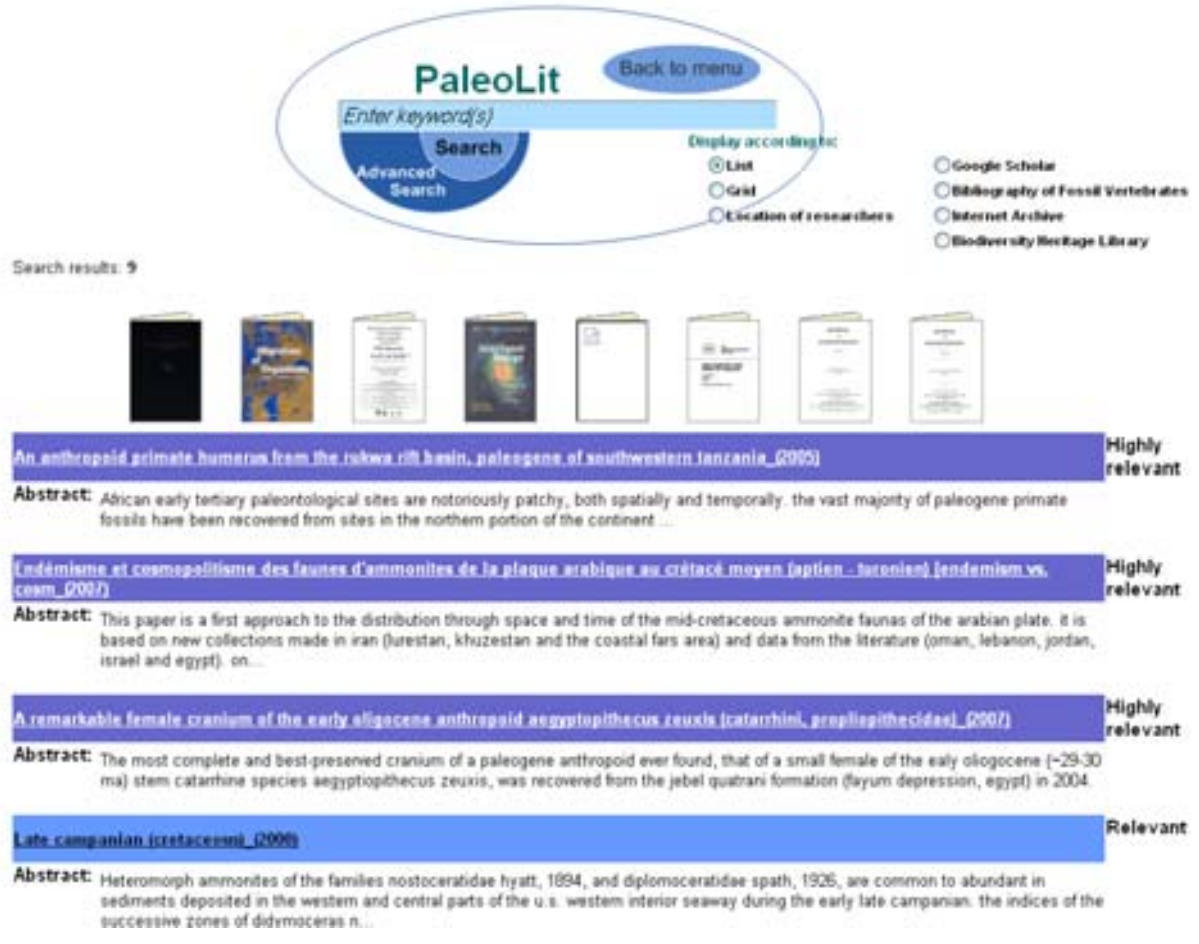

Fig. 8. PaleoLit version 2, list view, to show levels of uncertainty.

An individual's preferences as to results display have been known to change upon greater familiarity with the system. One of us involved in a Vivísimo user study found that users preferred the list view when they began to use Vivísimo, and after some use of the system, changed their preferences, believing that the document clusters provided a better overview of the document set as a whole. So while the participants in our PaleoLit study did not appreciate the horizontal result layout we call "grid view," we wonder whether they might prefer this layout with continued use. We could not run a longitudinal study with the current prototype since it holds insufficient articles to be useful for an extended period.

## 5 Principle Concerning Making Relevance Decisions

We believe the principle of time-optimization helps explain many of our findings. In the first study, participants showed that document title and abstract are important to making relevance decisions. This confirms that our system approach to ranking in the second study in which we consider bibliographic fields mirrors a human approach to relevance judgment. In the third study, many participants liked relevance labels, perhaps because the labels save time in result assessment.

We are not surprised that time optimization may explain judgments as to relevance since it has been found that time optimization is key in certain cognitive and physiological circumstances. Cognitively, it has been said that people in a well-known domain make decisions according to a series of learned steps [51]. In a domain with more unknowns, however, people apply weaker methods. One such method is to "satisfice" – a word coined to represent a mix of satisfy and sacrifice. Satisficing explains why people make good-enough decisions to save time rather than optimal decisions that tend to require more thought [51].

Physiologically, too, it has been found that humans forgo precision for the sake of time. Humans hold out an arm to catch a ball seen in air but adjust the arm position when the ball gets close. That is, humans make probabilistic, good-enough decisions and then adjust in time if found to be inaccurate.[14] It appears that neural activity builds until either a decision must be made, or more likely, until a threshold of activity is reached. Neurologists posit a trade-off between decision speed and accuracy which helps explain reaction time and error rate [13].

Accepting time optimization as a principle underlying our three preliminary studies guides our research. We should continue to compare user-derived and algorithm-derived uncertainty levels to improve correlation to the extent possible. We will also explore whether we have portrayed uncertainty levels optimally or whether we should add levels or even reduce from three levels to two, or give the user the option of selecting the number of relevance levels. We will consider also expanding the retrieval mechanism such that a query returns results from multiple sites and ranks them in a single list, in a federated search, so that the user does not need to visit each sites individually.

## 6 Summary Conclusion

We have tried to combine user perceptions of relevance with algorithm and interface design. We selected participants who have worked in paleontology to answer questions about article relevance and a prototype digital library, PaleoLit, that we created for the purpose.

We have conducted studies to determine how domain experts think about relevance, and how we might use this to model ranking for a search algorithm, and what we could provide visually to aid in relevance decisions. In a preliminary study with a small number of participants, we found high variance among what domain specialists consider relevance and to what degree those items are relevant, and that they were more likely to agree on what is not relevant that what is. According to some of our participants, the location of a key term in bibliographic fields in a journal article (title, abstract, caption, etc.) may be an indicator of which level of relevance the document should have. We included this in our ranking algorithm in addition to syntactic and lexical ontology-aided matching. Visually, we use labels and color blocks to indicate relevancy levels. While the value of our labels was acknowledged by a number of our participants in the third study, the horizontal layout was not, perhaps due to our choice of color, or to its unfamiliarity.

We connected our work to the principle that explains decision-making in terms of time optimization. Future researchers are welcome to test this further. Our most useful contribution may be our multi-disciplinary study of relevance that juxtaposes user relevance criteria, algorithm creation and testing, and results display within a particular domain.

## Appendix

The appendix lists bibliographic citations for the articles from the first study. They are numbered randomly,

[14] Robert T. Knight, "Oscillatory Dynamic and Brain Machine Interfaces," Keynote Address for Systems, Man, Cybernetics conference, October 12, 2009.

with the numbers corresponding to participant responses in Table 1.

**Dr. Judith Gelernter** is a postdoctoral fellow in the Language Technologies Institute of Carnegie Mellon University, with research projects also in the university's Institute for Software Research. She received a BA degree in Medieval culture from Yale University, an AM in Fine arts from Harvard University, and a PhD in Information science from Rutgers University. Her current research is in text mining, natural language processing, social network analysis and geoinformatics. She has published 12 articles and a book.

**Dong Cao** is doctoral student who has been visiting Carnegie Mellon University since 2008. He is enrolled at the Beijing University of Posts and Telecommunications in China where he expects to receive his PhD in computer science in 2010. He received his undergraduate degree at the China University of Mining and Technology in 2005. He has published four papers on semantic web services, ontology applications and document management. His current research interests include information retrieval and machine learning, especially in document management applications such as text classification and clustering.

**Dr. Jaime Carbonell** is the Director of the Language Technologies Institute and Allen Newell Professor of Computer Science at Carnegie Mellon University. He received SB degrees in Physics and Mathematics from MIT, and MS and PhD degrees in Computer Science from Yale University. His current research spans several areas, including virtually all aspects of Language Technologies: text mining, natural language processing, machine translation, and automated summarization (where he invented the MMR search-diversity method), question answering, etc. He is also an expert in Machine Learning, editing 3 books, and serving as editor-in-chief of the Machine Learning Journal for 4 years. Over all, he has published over 250 books and articles.